%% file: comp_imposs_res.tex
\documentclass[12pt, hidelinks]{article}
\usepackage[utf8]{inputenc}
\usepackage{amsmath, amssymb, url, float, graphicx}
\usepackage[small,bf]{caption}
\usepackage{fullpage}

\newcommand{\normsq}[1]{\left\|{#1}\right\|_2^2}
\newcommand{\tm}{\theta^\mathrm{max}}
\newcommand{\lagr}{\mathcal{L}}

\newcommand{\eps}{\varepsilon}
\newcommand{\complex}{\mathbf{C}}
\newcommand{\ii}{\mathbf{i}}

\input defs.tex

\title{Computational Bounds For Photonic Design}
\date{December 2018}
\author{Guillermo Angeris \and Jelena Vu\v{c}kovi\'c \and Stephen Boyd}

\begin{document}
\maketitle

\begin{abstract}
Physical design problems, such as photonic inverse design, are typically solved
using local optimization methods. These methods often produce what
appear to be good or very good designs when compared to classical
design methods, but it is not known how far from optimal such designs
really are.  We address this issue by developing methods for computing
a bound on the true optimal value of a physical design problem;
physical designs with objective smaller than our bound are impossible
to achieve.  Our bound is based on Lagrange duality and exploits the
special mathematical structure of these physical design problems.  For
a multi-mode 2D Helmholtz resonator, numerical examples show that the
bounds we compute are often close to the objective values obtained
using local optimization methods, which reveals that the designs are
not only good, but in fact nearly optimal.  Our computational bounding
method also produces, as a by-product, a reasonable starting point for
local optimization methods.
\end{abstract}

\section{Introduction}
Computer-aided design of physical systems is growing
rapidly in several fields, including photonics~\cite{molesky:2018} (where it is
known as inverse design),  horn design~\cite{noreland:2010}, and mechanical
design (aerospace, structures)~\cite{haftka:2012}. These design methods
formulate the physical design problem as a constrained nonconvex optimization
problem, and then use local optimization to attempt to solve the problem.
Commonly used methods include gradient descent, with adjoint-based evaluations
of the gradient~\cite{lalau:2013}, methods that alternate optimizing over the
structure and over the response~\cite{lu:2010}, and the alternating directions
method of multipliers (ADMM)~\cite{lu:2013}, among others. These methods can
be very effective, in the sense of producing what appear to be very good
physical designs, for example when compared to classical design approaches.

Because they are local optimization methods, they do not guarantee that a
globally optimal design is found, nor do we know how far from optimal the
resulting design is. This paper addresses the question of how far a physical
design is from globally optimal by computing a lower bound on the
optimal objective value of the optimization problem.  A lower bound on the
objective value can be interpreted as an impossibility result since it asserts
that no physical design can have a lower objective than a number we
compute.

Our bound is similar in spirit to analytical lower bounds, which give lower
bounds as simple formulas in terms of gross quantities like temperature and
wavelength, based on very simplified models and objectives, \eg, the Reynolds
number~\cite{purcell:1977}, the Carnot efficiency
limit~\cite[\S3.8]{fermi:1936}, or the optical diffraction
limit~\cite[\S8.6]{born:2013}. There has been some additional work in bounding some other quantities and figures of merit for optical systems, including the local density of states~\cite{miller:2016, shim:2018} for different types of materials, via fundamental physical principles. In contrast, our method computes a (numerical) lower bound for the optimization objective for each design problem.

In this paper, we derive a parametrized family of lower bounds on the optimal
objective for a class of physical design problems, using Lagrange duality.  We
can optimize over the parameter, to obtain the best (largest) lower bound, by
solving the Lagrange dual problem---which is convex even though the original
design problem is not. We illustrate our lower bound on a two-dimensional
multi-mode resonator. Our lower bound is close to the objective obtained by a
design using ADMM, which shows that the design, and indeed our lower bound, are
both very close to the global optimum.

\section{Physical design}
\subsection{Physical design problem}
In physical design, we design a structure so that the field, under a given
excitation,
is close to some desired or target field.  We parametrize the structure using a
vector $\theta$, and we denote the field by the vector $z$.
In photonic design, for example, we choose the index of refraction at each
rectangle on a grid, within limits, to achieve or get close to a desired 
electromagnetic field.

We can express this as the following optimization problem:
\begin{equation}
\begin{array}{ll}
\mbox{minimize}   & \frac12 \normsq{W(z - \hat z)} \\
\mbox{subject to} & (A+\diag(\theta))z=b \\
& 0 \le \theta\le \tm,
\end{array}
\label{eq:invdes}
\end{equation}
with variables $z \in \reals^n$ (the field) and $\theta \in \reals^n$, 
which describes the physical design.
The data are
the weight matrix $W \in \reals^{n\times n}$, which is diagonal with positive
diagonal entries, the desired or target field
$\hat z\in \reals^n$, the matrix $A \in \reals^{n \times n}$, the excitation
vector $b \in \reals^n$,
and the vector $\tm$ of limits on the physical design parameter $\theta$.
The constraint equation $(A +\diag (\theta))z = b$ encodes the physics of the
problem. We let $p^\star$ denote the optimal value of~(\ref{eq:invdes}).

We can handle the case when the lower limit on the 
physical parameter is nonzero, for example,
$\theta^\mathrm{min} \le \theta \le \tm$.  We do this by replacing the lower 
limit by $0$, the upper limit by $\tm - \theta^\mathrm{min}$, and replacing $A$
with $A+ \diag(\theta^\mathrm{min})$.
Additionally, the construction extends easily to the case where 
the field $z$, the matrix $A$, and the excitation $b$ have complex entries.

When the coefficient matrix in the physics equation 
$(A+\diag(\theta))z = b$ is nonsingular, there is a unique field,
$z = (A+\diag(\theta))^{-1}b$.
In some applications, however, the coefficient matrix is singular, 
and there is either no field that satisfies the equations, or many.
In the former case, we take the objective to be $+\infty$.
In the latter case, the set of solutions is an affine set and simple 
least squares can be used to find the field that satisfies the physics
equation and minimizes the objective.

An important special case occurs when we seek a mode (eigenvector) of
a system that is close to $\hat z$. To do this we take $b=0$ and 
subtract $\lambda I$ from the coefficient matrix, where $\lambda$ is the required 
eigenvalue.  We can handle the case of
unspecified eigenvalues by a simple extension described later 
in problem~(\ref{eq:undef-eigenvalue}), where $\lambda$ also becomes a design 
variable, subject to a lower and upper bound.

In the problem~(\ref{eq:invdes}), the physical design parameters
enter in a very specific way: as the diagonal entries of the coefficient matrix
of the physics equation.
Many physics equations have this form for a suitable definition of the field
$z$ and parameter $\theta$,
including the time-independent Sch\"odinger equation, Helmholtz's equation, the
heat equation, and Maxwell's equations in one dimension. (Maxwell's equations
in two and three dimensions are included in this formalism via the simple
extension given in problem~(\ref{eq:masks}).)

\paragraph{Boolean physical design problem.}
A variation on the problem~(\ref{eq:invdes}) replaces the 
physical parameter constraint $0 \leq \theta_j \leq \tm_j$ with 
the constraint $\theta_j \in \{0, \tm_j\}$, which limits each physical
parameter value to only two possible values.
(This occurs when we are choosing between two materials, such as 
silicon or air, in each of the patches in the structure we are designing.)
We refer to this modified problem as the \emph{Boolean physical design
problem},
as opposed to the continuous physical design problem~(\ref{eq:invdes}).
It is clear that the optimal value of the Boolean physical design
is no smaller than $p^\star$, the optimal value of the continuous physical 
design problem.

\subsection{Approximate solutions} The problem~(\ref{eq:invdes})
is not convex and generally hard to solve exactly~\cite{cvxbook}.
It is, however, bi-convex, since it is convex in $z$ when $\theta$ is fixed,
and convex in $\theta$ when $z$ is fixed. Using variations on this observation,
researchers have developed a number of methods for
approximately solving~(\ref{eq:invdes}) via heuristic means,
such as alternating
optimization over $z$ and $\theta$ on the augmented Lagrangian of this
problem~\cite{lu:2013}. 
Other heuristics can be used to find approximate solutions of the 
Boolean physical design problem.
These methods produce what appear to be very good physical designs when
compared to previous hand-crafted designs or classical designs.

\subsection{Performance bounds}
Since the approximate solution methods used are local and therefore 
heuristic, the question arises:
how far are these approximate designs from an optimal design?  In other words,
how far is the objective found by these methods from $p^\star$?
Suppose, for example, that a heuristic method finds a design with objective
value 13.1.
We do not know what the optimal objective $p^\star$ is, other than $p^\star
\leq 13.1$. Does there exist a design with objective value 10? Or 5?
Or are these values of the objective impossible, \ie, smaller than $p^\star$?

The method described in this paper aims to answer this question.
Specifically, we will compute a provable lower 
bound $L$ on the optimal objective value $p^\star$ of~(\ref{eq:invdes}).
In our example above, our method might compute the lower bound value $L=12.5$.
This means that no design can ever achieve an objective value smaller than
12.5.  It also means that 
a design with an objective value of 13.1 is not too far from optimal, 
since we would know that $L= 12.5 \leq p^\star \leq 13.1$.

A lower bound $L$ on $p^\star$ can be interpreted as an \emph{impossibility
result},
since it tells us that it is impossible for a physical design to achieve an
objective value less than $L$.
We can also interpret $L$ as a \emph{performance bound}.
The lower bound $L$ does not tell us what $p^\star$ is; it just gives a lower
limit on what it can be.
(An upper limit $U$ can be found by using any heuristic method, as the final 
objective value attained.)

We note that the lower bound $L$ we find on $p^\star$ also serves as a lower
bound on the optimal value of the Boolean physical design problem, since its
optimal value is larger than or equal to $p^\star$.

\section{Performance bounds via Lagrange duality}\label{sec:bounds}
In this section, we explain our lower bound method.

\subsection{Lagrangian duality}
We first rewrite~(\ref{eq:invdes}) as
\begin{equation}\label{eq:ind-problem}
\begin{array}{ll}
\mbox{minimize}   & \frac12 \normsq{W(z - \hat z)} + I(\theta)\\
\mbox{subject to} & (A+\diag(\theta))z=b,
\end{array}
\end{equation}
where $I$ is an indicator function, \ie, $I(\theta)=0$
when $0 \le \theta \le \tm$ and $+\infty$ otherwise.
The Lagrangian of this problem is
\begin{equation}\label{eq:lagrangian}
  \lagr(z, \theta, \nu) = \frac12\normsq{W(z - \hat z)} + I(\theta)
+ \nu^T((A + \diag(\theta))z-b),
\end{equation}
where $\nu\in\reals^n$ is a dual variable. The Lagrange dual function is
\[
g(\nu) = \inf_{\theta, z} \lagr(z, \theta, \nu).
\]
(See~\cite[Chapter 5]{cvxbook}.)
It is a basic and easily proved fact that for any $\nu$, we have 
$g(\nu) \leq p^\star$ (see~\cite[\S5.1.3]{cvxbook}).
In other words, $g(\nu)$ is a lower bound on $p^\star$.
While $g(\nu)$ always gives a lower bound on $p^\star$, the challenge for
nonconvex problems such as~(\ref{eq:invdes}) is to \emph{evaluate} $g(\nu)$.  
We will see now that this can be done for our problem~(\ref{eq:invdes}).

\subsection{Evaluating the dual function}
To evaluate $g(\nu)$ we must minimize $\lagr(z,\theta,\nu)$ over $z$ and
$\theta$.
Since for each $\theta$, $\lagr(z,\theta,\nu)$ is convex quadratic in $z$,
we can analytically carry out the minimization over $z$.  We have
\begin{align}
g(\nu) &= \inf_\theta \inf_{z} \lagr(z, \theta, \nu)\nonumber\\
&= \inf_\theta  \left( -\frac{1}{2} \normsq{W^{-1}((A+\diag(\theta))^T\nu - W^2
\hat z)} -
\nu^Tb + \frac12\normsq{W\hat z} + I(\theta)\right)\nonumber\\
&= \inf_{0 \le \theta\le \tm} -\frac{1}{2}
\normsq{W^{-1}((A+\diag(\theta))^T\nu
- W^2 \hat z)} - \nu^Tb + \frac12\normsq{W\hat z}
\label{eq:theta-max}.
\end{align}
We can see that this is true since the minimizer of the only terms depending on
$z$,
\[
\argmin_z\left(\frac12 \normsq{W(z - \hat z)} + \nu^T (A +
\diag(\theta))z\right),
\]
can be found by taking the gradient and setting it to zero (which is necessary
and sufficient by convexity and differentiability). This gives that the
minimizing
$z$ is
\begin{equation}\label{eq:argmin-z}
z = \hat z - W^{-2}(A + \diag(\theta))\nu,
\end{equation}
which yields~(\ref{eq:theta-max}) when plugged in.

The expression in~(\ref{eq:theta-max}) is separable over each $\theta_i$; it
can be rewritten as
\begin{align}
g(\nu) &= 
\inf_{0 \le \theta\le \tm} -\frac{1}{2} \sum_{j=1}^n
W_{jj}^{-2}\left((A^T\nu)_j + \nu_j\theta_j - W^2_{jj} \hat z_j\right)^2 -
\nu^Tb + \frac12\normsq{W\hat z} \nonumber \\
&= 
\sum_{j=1}^n\left(\inf_{0 \le \theta_j\le \tm_j} -\frac{1}{2}
W_{jj}^{-2}\left((A^T\nu)_j + \nu_j\theta_j - W^2_{jj} \hat z_j\right)^2\right)
-
\nu^Tb + \frac12\normsq{W\hat z} \nonumber \\
&= -\frac12 \sum_{j=1}^n W_{jj}^{-2}\max\left\{\left(a^T_j\nu - W^2_{jj} \hat
z_j\right)^2, \left(a^T_j\nu + \nu_j\tm_j - W^2_{jj} \hat z_j\right)^2\right\}
- \nu^Tb + \frac12 \normsq{W\hat z}, \label{eq:dual-function}
\end{align}
where $a_j$ is the $j$th column of $A$.
In the last line, we use the basic fact that a scalar convex quadratic function
achieves its maximum over an interval at the interval's boundary.

With this simple expression for the dual function, we can now generate
lower bounds on $p^\star$, by simply evaluating it for any $\nu$.
We note that $g$ is also the dual function of the Boolean
physical design problem.

\subsection{Dual optimization problem}
It is natural to seek the best or largest lower bound on $p^\star$, by 
choosing $\nu$ that maximizes our lower bound.  This leads to the 
dual problem (see~\cite[\S5.2]{cvxbook}),
\[
\begin{array}{ll}
\mbox{maximize} & g(\nu),
\end{array}
\]
with variable $\nu$.
We denote the optimal value as $d^\star$, which is the best lower bound
on $p^\star$ that can be found from the Lagrange dual function.
The dual problem is always a convex optimization problem
(see~\cite[\S5.1.2]{cvxbook}); to effectively use it, we need a way
to tractably maximize $g$, which we have in our case, since the
dual problem can be expressed as
the convex quadratically-constrained quadratic program (QCQP)
\begin{equation}
\begin{array}{ll}
\mbox{maximize}   & -(1/2)\ones^Tt - \nu^Tb + (1/2) \normsq{W\hat z} \\
\mbox{subject to} & t_j \ge W_{jj}^{-2}\left(a_j^T\nu - W_{jj}^2 \hat
z_j\right)^2, ~~ j=1, \dots, n\\
& t_j \ge W_{jj}^{-2}\left(a_j^T\nu +\nu_j\tm_j - W_{jj}^2\hat z_j\right)^2, ~~
j=1, \dots, n,
\end{array}
\label{eq:dual}
\end{equation}
with variables $t$ and $\nu$. This problem is easily solved and its optimal
value, $d^\star$, is a lower bound on $p^\star$.

The dual optimization problem~(\ref{eq:dual}) can be solved 
several ways, including via ADMM (which can exploit the fact 
that all subproblems are quadratic; see~\cite{boyd:2011}),
interior point methods (see~\cite[\S11.1]{cvxbook}), or by 
rewriting it as a second-order cone program (SOCP) (see~\cite{lobo:1998}; this can also be done automatically by modeling languages such as CVXPY~\cite{cvxpy:2018})
and then using one of the many available SOCP solvers, such as 
SCS~\cite{scs_paper:2016, scs:2016}, ECOS~\cite{ecos:2013}, or Gurobi~\cite{gurobi:2018}.
We also note that the dual problem does not have to be perfectly solved;
we get a lower bound for \emph{any} value of the dual variable $\nu$.

In this paper, we used the Gurobi solver to solve a (sparse) program with $n=63001$, which took approximately 8 minutes to solve on a two-core Intel Core i5 machine with 8GB of RAM. By further exploiting the structure of the problem, giving good initializations, or by using less accurate methods when small tolerances are not required, it is likely that these problems could be solved even more quickly, for larger systems.

\subsection{Initializations via Lagrange dual}\label{sec:initalization}
The solution of the Lagrange dual problem can be used to suggest starting
points
in a heuristic or local method for approximately solving~(\ref{eq:invdes}).

\paragraph{Initial structure.}
Let $\nu^\star$ be a solution of the dual problem~(\ref{eq:dual}).
We can take as initial structure $\theta^0$ which
minimizes~(\ref{eq:theta-max}), \ie,
\[
\theta^0_j \in \argmax_{\theta_j \in \{0, \,\tm_j\}} \left(a_j^T\nu^\star
+\nu_j^\star\theta_j - W_{jj}^2\hat z_j\right)^2.
\]
This choice of initial structure is feasible for~(\ref{eq:invdes}) and, in
fact, is feasible for the Boolean physical design problem as well.

\paragraph{Initial field.}
One way to obtain an initial field is to simply solve
the physics equation for $\theta^0$, when the physics coefficient matrix
is nonsingular.  When it is singular, but the physics equation is solvable,
we compute $z$ as the field that minimizes the objective, subject to the 
physics equation.
This gives a feasible field, but in some cases the resulting point
is not very useful.
For example when $b=0$, and the coefficient matrix is nonsingular, 
we obtain $z^0 = 0$. 

Another possibility is to find the minimizer of the Lagrangian with
the given structure and an optimal dual variable value, \ie,
\[
z^0 = \argmin_{z} \lagr(\theta^0, z, \nu^\star).
\]
The value is already given in~(\ref{eq:argmin-z}):
\[
z^0 = \hat z - W^{-2}\left(A + \diag(\theta^0)\right)^T\nu^\star.
\]
This initial field is not feasible, \ie, it does not satisfy the physics
equation, but it seems to be a very good initial choice for heuristic
algorithms.

\section{Multi-scenario design}\label{sec:multi-scenario}
In this section we mention an extension of our basic problem~(\ref{eq:invdes}),
in which we wish to design one physical structure that gives reasonable 
performance in $N$ different scenarios.
The scenarios can represent different operating temperatures, different 
frequencies, or different modes of excitation.

We will index the scenarios by the superscript $i$, with $i=1, \ldots, N$.
Each scenario can have a different weight matrix $W^i$, a different target
field $\hat z^i$, a different physics matrix $A^i$, and a different
excitation $b^i$. We have only one physical design variable $\theta$,
and $N$ different field responses, $z^i$, $i=1, \ldots, N$.
We take as our overall objective the sum (or average) of the objectives
under the scenarios.  This leads to the problem
\begin{equation}\label{eq:invdes-multi}
\begin{array}{ll}
\mbox{minimize}   & \frac12\sum_{i=1}^N \normsq{W^i(z^i - \hat z^i)}\\
\mbox{subject to} & (A^i + \diag(\theta))z^i =b^i,~~i=1,\dots,N \\
& 0 \le \theta \le \tm,
\end{array}
\end{equation}
with variables $\theta$ (the structure) and $z^i$ (the fields under the $N$
different
scenarios).

Our bounding method easily generalizes to this multi-scenario physical design
problem.

\paragraph{Dual optimization problem.} As before, define $a^i_j$ to be the
$j$th column of $A^i$ and allow
$\nu^i$ to be the Lagrange multiplier for the $i$th constraint, then the new
dual problem is,
\begin{equation}\label{eq:complete-multi}
\begin{array}{ll}
\mbox{maximize}   & -(1/2)\ones^Tt - \sum_{i=1}^N (\nu^i)^T(b^i) + (1/2)
\sum_{i=1}^N \normsq{W^i\hat z^i} \\
\mbox{subject to} & t_j \ge \sum_{i=1}^N (W_{jj}^i)^{-2}\left((a_j^i)^T\nu^i -
(W_{jj}^i)^2\hat z_j^i\right)^2, ~~ j=1, \dots, n\\
& t_j \ge \sum_{i=1}^N (W_{jj}^i)^{-2}\left((a_j^i)^T\nu^i
+\nu_j^i\tm_j - (W_{jj}^i)^2\hat z_j^i\right)^2, ~~ j=1, \dots, n,\\
\end{array}
\end{equation}
which is also a convex QCQP. This new dual optimization problem can be derived
in a similar way to the construction of \S\ref{sec:bounds}.

\paragraph{Initial structure and fields.} Similar initializations hold
for~(\ref{eq:invdes-multi}) as do for~(\ref{eq:invdes}). We can find
an initial $\theta^0$ given by 
\begin{equation}\label{eq:theta-init}
\theta^0_j \in \argmax_{\theta_j \in \{0, \,\tm_j\}} \left(\sum_i
(W_{jj}^i)^{-2}\left((a_j^i)^T(\nu^i)^\star +(\nu^i)^\star_j\theta_j -
(W_{jj}^i)^2\hat z_j^i\right)^2\right),
\end{equation}
while we can find feasible initial fields by solving the physics equations for
each scenario, or as the minimizer of the Lagrangian,
\begin{equation}\label{eq:field-init}
(z^i)^0 = \hat z^i - (W^i)^{-2}\left(A^i +
\diag(\theta^0)\right)^T(\nu^i)^\star,
\end{equation}
for $i=1,\dots,N$, which gives infeasible fields (often, however, these fields are good 
initializations).

\section{Numerical example}

\subsection{Physics and discretization}
We begin with Helmholtz's equation in two dimensions,
\begin{equation}\label{eq:helmholtz}
\nabla^2f(x, y) + \left(\frac{\omega}{c(x, y)}\right)^2f(x, y) = 0,
\end{equation}
where $f: \reals^2 \to \reals$ is a function representing the wave's amplitude,
$\nabla^2= \partial_x^2 + \partial_y^2$ is the Laplacian in two dimensions,
$\omega \in \reals_+$ is the angular frequency of the wave, and $c:
\reals^2 \to \reals_+$ is the speed of the wave in the material at position
$(x, y)$, which we can change by an appropriate choice of material. For this
problem, we will allow the choice of any material that has a propagation speed
between $0 < c^\mathrm{min}(x, y) \le c(x, y) \le c^\mathrm{max}(x, y)$, such
that $f$ is close to $\hat f$, some desired field.

Throughout, we will also assume Dirichlet boundary conditions for convenience
(that is, $f(x, y) = 0$, whenever $(x, y)$ is on the boundary of the domain),
though any other boundary conditions could be similarly used with this
method.

We discretize each of $c$, $f$, and $\nabla^2$
in equation~(\ref{eq:helmholtz}) using a simple
finite-difference approximation over an equally-spaced
rectilinear grid.  (More sophisticated discretization methods would 
also work with our method.)
Specifically, let ${(x_i, y_i)}$ for $i=1, \dots, n$ be the discretized
points of the grid, with separation distance $h$ (\eg, $y_{i+1} - y_i = x_{i+1}
- x_i = h$). We then let $z$ and $\hat z$, both in $\reals^n$, be the
discretization of $f$ and $\hat f$, respectively, over the grid,
\[
z_i = f(x_i, y_i), ~~ \hat z_i = \hat f(x_i, y_i).
\]
Using this discretization, we can approximate the second derivative of $f$ at
the grid points as,
\[
\partial_x^2 f(x_i, y_i) \approx \frac{f(x_i+h, y_i) - 2f(x_i, y_i) + f(x_i -
h, y_i)}{h^2} = \Delta_x z,
\]
for some matrix $\Delta_x$, and similarly for $\partial_y^2$, whose finite
approximation we will call $\Delta_y$. We can then define a complete
approximate Laplacian as the sum of the two matrices,
\[
\Delta = \Delta_x + \Delta_y.
\]
We also similarly discretize $c(x, y)$ as
\[
\theta_i = \frac{1}{c(x_i, y_i)^2},
\]
where $\theta \in \reals^n$. The constraints on $c(x, y)$ become
\[
\theta^\mathrm{min} = \frac{1}{c^\mathrm{max}(x_i, y_i)^2} \le \theta_i \le
\frac{1}{c^\mathrm{min}(x_i, y_i)^2} = \tm.
\vspace{.5em} 
\]

We can now write the fully-discretized form of Helmholtz's
equation as
\[
(\Delta + \omega^2\diag(\theta))z=0
\]
or, equivalently,
\[
\left(\frac{1}{\omega^2}\Delta + \diag(\theta)\right) z=0.
\]
So the final problem is, after replacing $\theta$ with $\theta -
\theta^\mathrm{min}$,
\[
\begin{array}{ll}
\mbox{minimize}   & \frac12 \normsq{W(z - \hat z)} \\
\mbox{subject to} & \left((1/\omega^2)\Delta +
\diag(\theta^\mathrm{min})+\diag(\theta)\right)z=0 \\
& 0 \le \theta\le \tm - \theta^\mathrm{min}.
\end{array}
\]
This has the form of problem~(\ref{eq:invdes}), with
\[
A = \frac{1}{\omega^2}\Delta + \diag(\theta^\mathrm{min}), \quad b = 0.
\]
Note that the design we are looking for---one that supports non-vanishing modes at each frequency---will, in general, have a singular (or indeterminate) physics equation. More specifically, the final design's physics equations will each have a linear set of solutions, from which we pick the one that minimizes the least squares residual in the objective.

\subsection{Problem data}\label{sec:problem-data}
\begin{figure}
\begin{center}
    \includegraphics[width=.98\textwidth]{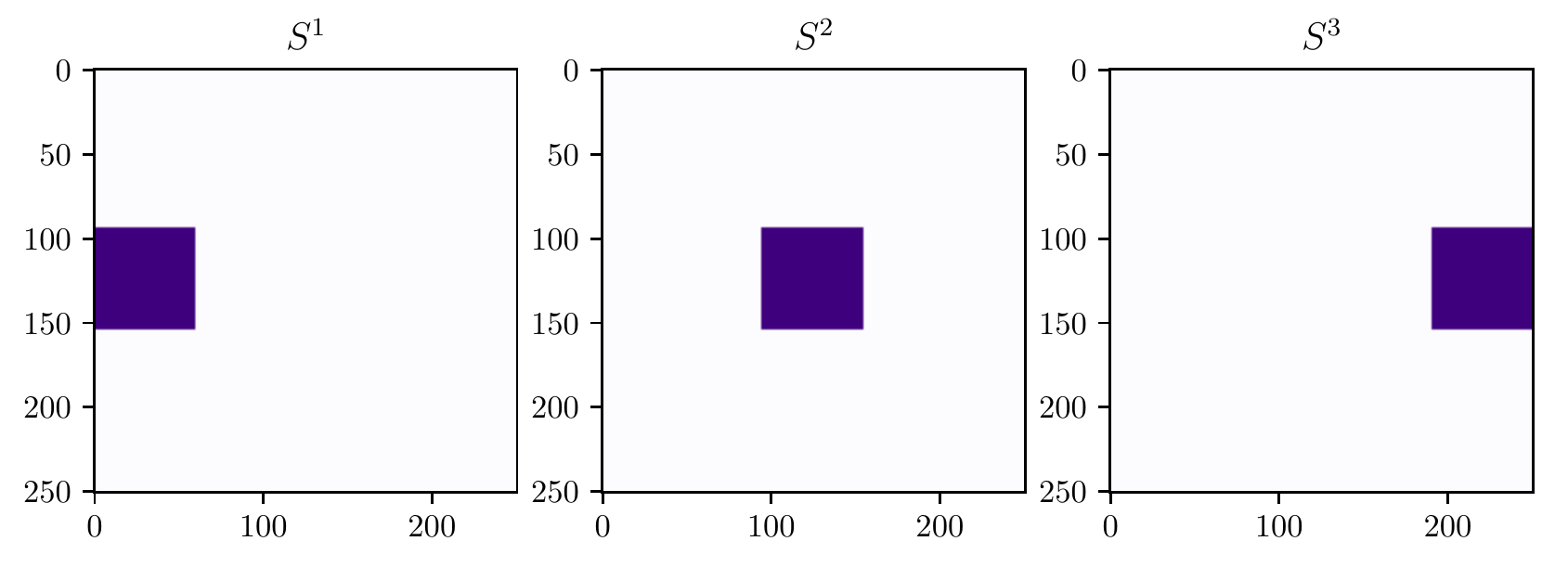}
\end{center}
\caption{The three target resonator regions.}
\label{fig:boxes}
\end{figure}

In this example, we will design a 2D resonator with modes that
are localized in the boxes found in figure~\ref{fig:boxes}, 
at each of three specified frequencies.
More specifically, let $S^i$ be the indices at frequency $i$ corresponding to
the boxes shown in figure~\ref{fig:boxes}.
We define the target field for frequency $i$ as
\[
\hat z^i_j = \begin{cases}
  1, & j \in S^i\\
  0, & j \not\in S^i.
\end{cases}
\]
We set the weights within the box containing the mode to be one and set
those outside the box to be larger:
\[
W^i_{jj} = \begin{cases}
  1, & j \in S^i\\
  5, & j \not\in S^i.
\end{cases}
\]

We specify three frequencies (\ie, $N=3$),
\[
\omega = (30\pi, 40\pi, 50\pi),
\]
at which to generate the specified modes by picking the propagation speed of the wave at each discretization point of the domain. We constrain the allowed propagation speed by picking
\[
\theta^\mathrm{min}_j = 1, \quad \tm_j = 2, \quad j = 1, \dots, n.
\]
Our discretization uses a $251 \times 251$ grid, so $n=251^2 = 63001$, with
$h=1/n$.

\subsection{Physical design}\label{sec:approx-minimization}
We use ADMM to approximately solve the physical design problem, as
in~\cite{lu:2013}, using penalty parameter $\rho =100$.
We initialized the method using the feasible structure and fields 
from~\S\ref{sec:numerical-dual}, though similar designs are achieved with 
simple initializations like $\theta = \theta^\mathrm{min}$ and $z^i = 0$, 
for $i=1, 2, 3$.
We stop the algorithm when the physics constraint residual norm drops 
below a fixed tolerance of $10^{-2}$.
The resulting locally optimized design is shown in figure~\ref{fig:resonator-theta} and 
the associated fields are shown in figure~\ref{fig:resonator-z}. 
In particular, after local optimization, we receive some $\theta$ and $z$ with
\[
\theta^\mathrm{min} \le \theta \le \tm, ~~ \left\|(A+\diag(\theta))z - b\right\|_2 \le 10^{-2},
\]
and then evaluate
\[
p = \frac12 \sum_{i=1}^3\normsq{W^i(z^i - \hat z^i)},
\]
which gives $p = 5145$.

Our non-optimized implementation required around 1.5 seconds per iteration
and took 332 iterations to converge to the specified tolerance, so the total physical design time is a bit under 9 minutes on a 2015 2.9GHz dual core
MacBook Pro. Our implementation used a sparse-direct solver; an iterative
CG solver with warm-start would have been much faster.

\subsection{Dual problem}\label{sec:numerical-dual}
We solved problem~(\ref{eq:complete-multi}) using the Gurobi~\cite{gurobi:2018} 
SOCP solver and the JuMP~\cite{dunning:2017} mathematical modeling language for 
Julia~\cite{bezanson:2017}.
Gurobi required under ten minutes to solve the dual problem, about the same
time required by the physical design.
This time, too, could be very much shortened; for example, we do not need to solve
the dual problem to the high accuracy that Gurobi delivers.

\begin{figure}
\begin{center}
    \includegraphics[width=.9\textwidth]{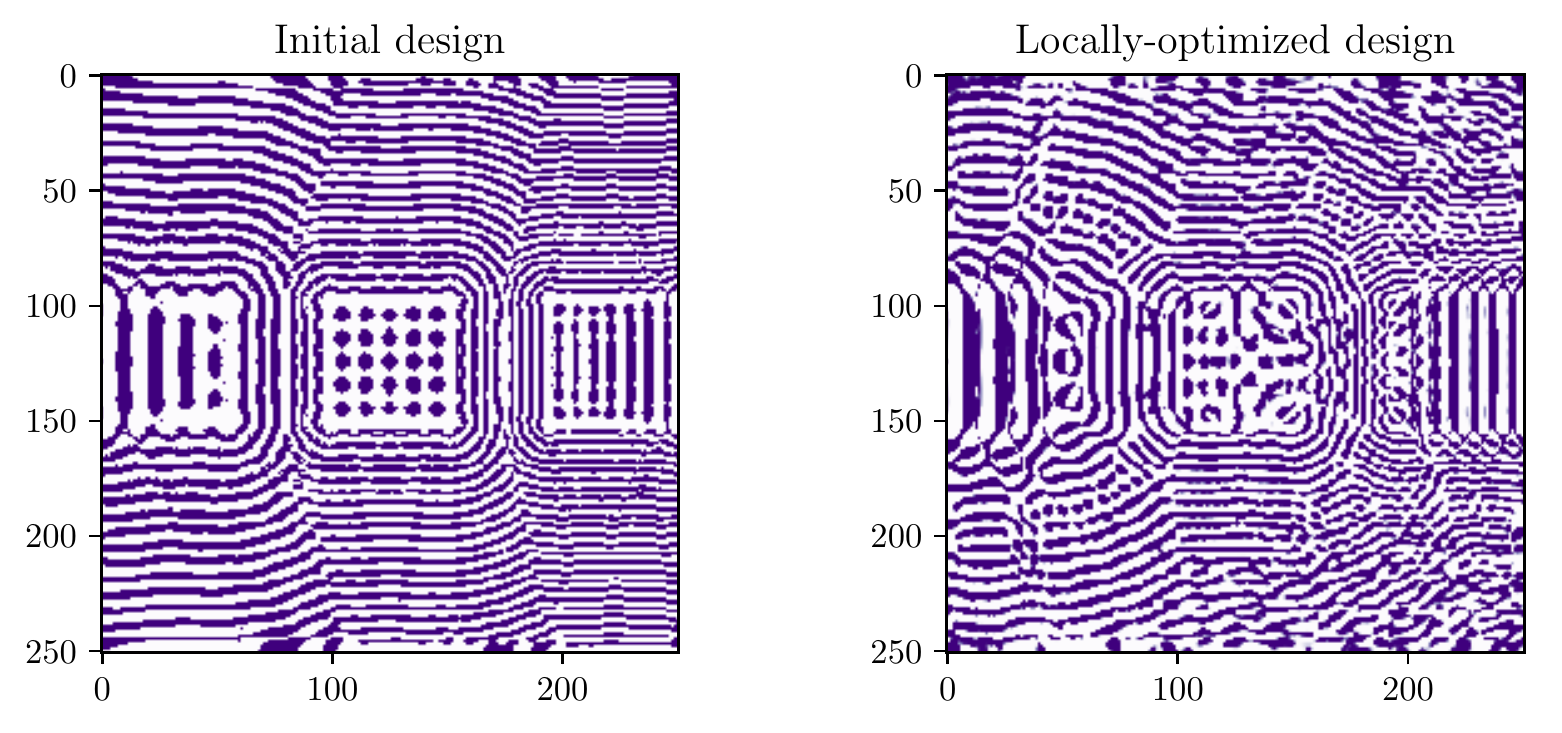}
\end{center}
\caption{\emph{Left.} Initial design suggested by the dual solution.
\emph{Right.} Optimized physical design.}
\label{fig:resonator-theta}
\end{figure}

\begin{figure}
\begin{center}
    \includegraphics[width=.98\textwidth]{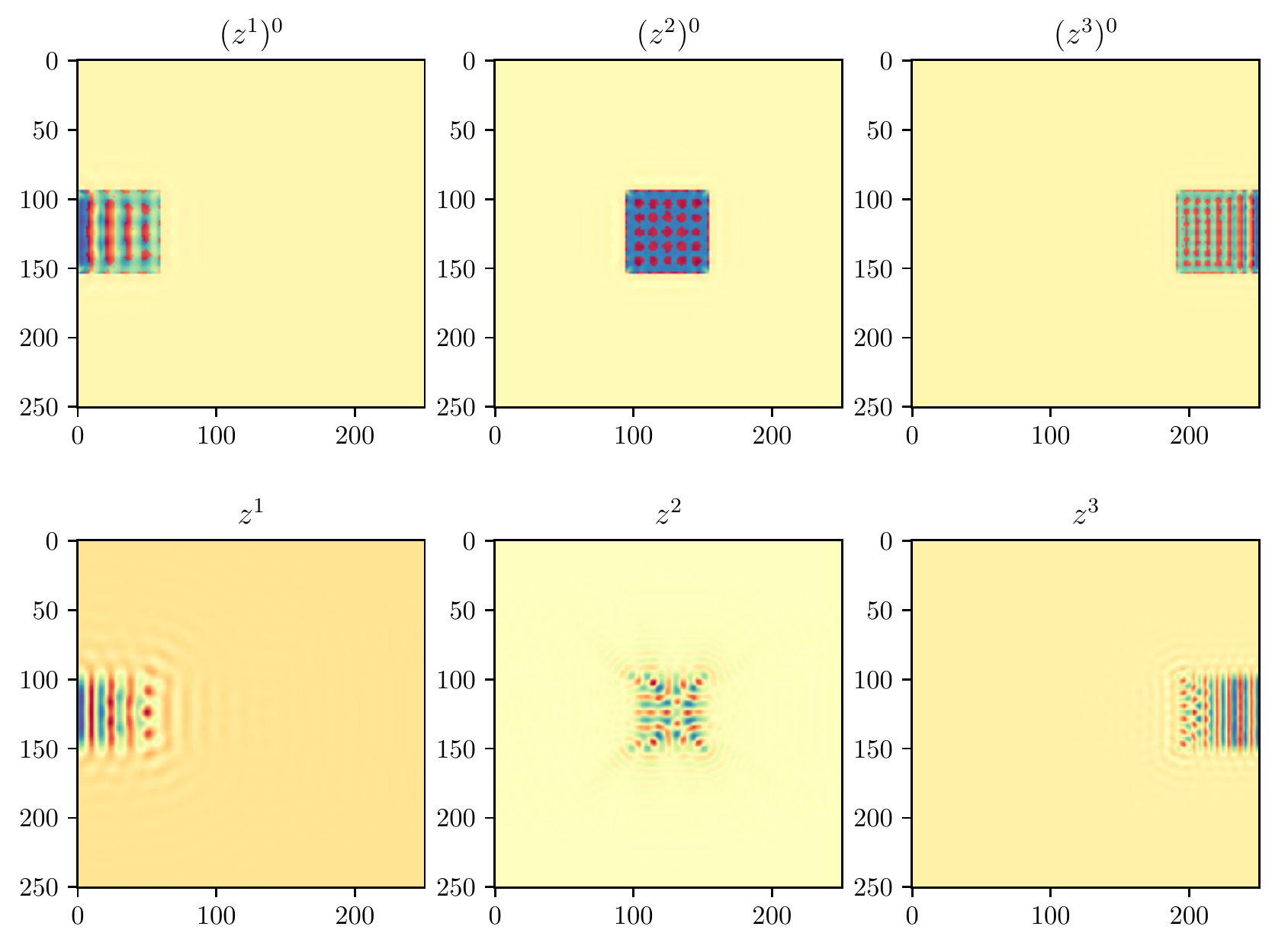}
\end{center}
\caption{\emph{Top row.} Fields suggested by solution to the dual problem. 
\emph{Bottom row.} Fields in ADMM physical design. Columns show the three frequencies.}
\label{fig:resonator-z}
\end{figure}

The optimal dual value found is $d^\star = 4733$,
with the initial design and fields suggested by the optimal dual solution shown
in figure~\ref{fig:resonator-theta} and figure~\ref{fig:resonator-z}, respectively. 
This tells us that
\[
4733 = d^\star \le p^\star \le p = 5145,
\]
which implies that our physical design objective value is no more than 
$(5145-4733)/4733 \approx 8.7\%$ suboptimal.
(We strongly suspect that $p^\star$ is closer to our design's value, 5145, than 
the lower bound, 4733.)

\section{Further extensions}
There are several straightforward extensions of the above problem, which may
yield useful results in specific circumstances. All of these problems have
analytic forms for their Lagrange dual functions, and all forms generalize
easily to their multi-frequency counterparts. Additionally, we explicitly derive the dual functions for some extensions which require a little more care.

\paragraph{Equality-constrained parameters.}
Sometimes, it might be the case that a single design parameter might control
several points in the domain of $z$---for example, in the case of Maxwell's
equations in two and three dimensions (see the appendix for more details), or when the domain's grid size is much smaller than the smallest features that can be constructed.

Let $S_k \subseteq \{1, \dots, n\}$ for $k=1, \dots, m$ be a partition of
indices, $\{1, \dots, n\}$. In other words, we want $S_k$ for $k=1, \dots, m$
to satisfy,
\[
\bigcup_{k=1}^m S_k = \{1, \dots, n\}
\]
and $S_k \cap S_l = \emptyset$ whenever $k \ne l$. These sets $S_k$ will
indicate the sets of indices which are constrained to be equal---conversely,
indices that are not constrained to be equal to any other indices are
represented by singleton sets.

We can then write the new optimization problem as
\begin{equation}\label{eq:masks}
\begin{array}{ll}
\mbox{minimize}   & \frac12 \normsq{W(z - \hat z)} \\
\mbox{subject to} & (A+\diag(\theta))z=b \\
& \theta_i = \theta_j, ~ \text{for all}~~ i,j \in S_k,~~ k = 1, \dots, m\\
& 0 \le \theta\le \tm.
\end{array}
\end{equation}

To compute the Lagrange dual, let $I'$ be an indicator function with $I'(\theta) = 0$ whenever $0 \le \theta \le \tm$ and $\theta_i = \theta_j$ for all $i,j \in S_k$ for $k=1, \dots, m$. Otherwise, $I'(\theta) = +\infty$. We can write the new problem as problem~(\ref{eq:ind-problem}) with the same Lagrangian as the one given in~(\ref{eq:lagrangian}), replacing $I$ with $I'$ in both expressions.

Minimization over $z$ is identical to~(\ref{eq:theta-max}) and minimization over $\theta$ is similar minus the fact that for each $k$, the indices found in $S_{k}$ are all constrained to be equal. Since the sum of convex quadratics is still a convex quadratic and, as before, since convex quadratics achieve minima at the boundary of an interval, we have
\begin{align*}
g(\nu) = -\frac12 \sum_{k=1}^m \max\Bigg\{\sum_{j \in S_k}W_{jj}^{-2}\left(a^T_j\nu - W^2_{jj} \hat
z_j\right)^2, \sum_{j \in S_k}W_{jj}^{-2}\big(a^T_j\nu + \nu_j\tm_j - &W^2_{jj} \hat z_j\big)^2\Bigg\}\\
&- \nu^Tb + \frac12 \normsq{W\hat z},
\end{align*}
as the final Lagrange dual function. The corresponding dual problem can be written as a convex QCQP.

\paragraph{Field constraints.}
In the case where~(\ref{eq:invdes}) has field constraints, \ie,
\[
\begin{array}{ll}
\mbox{minimize}   & \frac12 \normsq{W(z - \hat z)} \\
\mbox{subject to} & (A+\diag(\theta))z=b \\
& (z_j - h_j)^2 \le (z^\mathrm{max}_j)^2, ~~ j=1,\dots, n\\
& 0 \le \theta\le \tm,
\end{array}
\]
for some $h\in \reals^n$, the construction also parallels the one given in
\S\ref{sec:bounds}. The resulting dual optimization problem, in
comparison to problem~(\ref{eq:dual}), cannot be written as a QCQP---it is, instead, a more general SOCP.

\paragraph{Regularizers.}
It is also possible to add a separable regularization term for $\theta$, the
parametrization of the device; for example, in the case where we would want to
bias specific $\theta_j$ towards either $0$ or $\tm_j$.

If we have a family of concave functions, $r_j:\reals \to \reals$ such that our
regularizer can be written as a function of the form
\[
\theta \mapsto \sum_{j=1}^n r_j(\theta_j),
\]
(one such example is a linear function of $\theta$), then the problem becomes
\[
\begin{array}{ll}
\mbox{minimize}   & \frac12 \normsq{W(z - \hat z)} + \sum_{j=1}^n r_j(\theta_j)
\\
\mbox{subject to} & (A+\diag(\theta))z=b \\
& 0 \le \theta\le \tm.
\end{array}
\]
By using the fact that $r_j$ is concave and therefore achieves a minimum over
an interval at the boundary of the interval, it is possible to derive a bound
that parallels~(\ref{eq:dual-function}).

\paragraph{Parameter perturbations.}
In some cases (\eg, when considering temperature perturbations), it might be
very natural to have a physical constraint of the form
\[
(A + D\diag(\theta))z = b,
\]
where $D \in \reals^{n\times n}$ is a diagonal matrix that is not necessarily invertible. In other words, our new problem is
\[
\begin{array}{ll}
\mbox{minimize}   & \frac12 \normsq{W(z - \hat z)} \\
\mbox{subject to} & (A+D\diag(\theta))z=b \\
& 0 \le \theta\le \tm.
\end{array}
\]
Directly applying the method from \S\ref{sec:bounds} yields a similar
explicit form for $g$ as given in~(\ref{eq:dual-function}).

\paragraph{Indeterminate eigenvalue.}
In the case where we want $z$ to be a mode of the device with some
unspecified eigenvalue $\lambda$ with upper and lower limits
$\lambda^\mathrm{min} \le \lambda \le \lambda^\mathrm{max}$, we can write the
problem as
\begin{equation}\label{eq:undef-eigenvalue}
\begin{array}{ll}
\mbox{minimize}   & \frac12 \normsq{W(z - \hat z)} \\
\mbox{subject to} & (A + \lambda I + \diag(\theta))z=b \\
& \lambda^\mathrm{min} \le \lambda \le \lambda^\mathrm{max}\\
& 0 \le \theta\le \tm.
\end{array}
\end{equation}

To construct the dual, note that the Lagrangian of this problem is similar to the Lagrangian of problem~(\ref{eq:invdes}),
\[
\lagr(z, \theta, \lambda, \nu) = \frac12\normsq{W(z - \hat z)} + I(\theta)
+ \nu^T((A + \lambda I + \diag(\theta))z-b).
\]
We will define the partial Lagrangian, $\lagr^p$ to be the infimum of $\lagr$ with respect to $z$ and $\theta$, leaving $\lambda$ and $\nu$ as free variables. The solution to the partial minimization of $\lagr$ is given in~(\ref{eq:dual-function}),
\begin{align*}
\lagr^p(\lambda, \nu) &= \inf_{z, \theta}\lagr(z, \theta, \lambda, \nu) \\
&= -\frac12 \sum_{j=1}^n W_{jj}^{-2}\max_{\theta_j \in \{0, \tm_j\}}\left(a^T_j\nu + (\lambda + \theta_j)\nu_j- W^2_{jj} \hat z_j\right)^2 - \nu^Tb + \frac12 \normsq{W\hat z}.
\end{align*}
As $\lagr^p(\lambda, \nu)$ is a concave in $\lambda$, it achieves its minimum at the boundaries of the domain of $\lambda$. So, since
\[
g(\nu) = \inf_{\lambda^\mathrm{min} \le \lambda \le \lambda^\mathrm{max}} \lagr^p(\lambda, \nu),
\]
we can write,
\[
g(\nu) = \min_{\lambda \in \{\lambda^\mathrm{min},\,\lambda^\mathrm{max}\}} \lagr^p(\lambda, \nu)
\]
which is the minimum over a (finite) number of concave functions. The corresponding dual problem can then be expressed as a convex QCQP.

\section{Conclusion}
This paper has derived a set of lower bounds for a general class of physical
design
problems, making it possible to give (a) an easily-computable certificate that
certain objectives cannot be physically achieved and (b) a bound on how
suboptimal (relative to the global optimum) a given design could be. Additionally,
as a side-effect of computing this lower bound, we also receive an
initialization for any heuristic approach we might take for approximately
solving~(\ref{eq:invdes}) or its multi-frequency
version~(\ref{eq:invdes-multi}).

Additionally, it seems feasible to obtain asymptotic bounds with respect to
physical parameters (\eg, with respect to the size of the device) via this
approach, since the optimization problem in~(\ref{eq:dual}) can easily be
written in an unconstrained form. In other words, picking any $\nu \in \reals^n$
will yield \emph{some} lower bound, and an appropriate choice might yield
scaling laws that could be useful as general rules-of-thumb in inverse design.

\section*{Acknowledgements}
We thank the Gordon and Betty Moore Foundation and Google for financial support. The authors would also like to thank Rahul Trivedi and Logan Su for useful discussions and help with debugging both code and derivations.

\newpage

\bibliography{citations}

\newpage
\section{Appendix}
\subsection{Optimization using ADMM}
We can approximately minimize~(\ref{eq:invdes}) via the alternating direction method of multipliers, as in~\cite{lu:2013}. The method proceeds by forming the augmented Lagrangian of~(\ref{eq:invdes}) and minimizing over each available variable, before updating a dual variable after each iteration.

\paragraph{ADMM iteration.}
We form the augmented lagrangian of problem~(\ref{eq:invdes}) as in~\cite{admm:2011}.
\[
\lagr^\mathrm{aug}(z, \theta; \nu) = \frac12\normsq{W(z - \hat z)} + \frac{\rho}{2}\normsq{(A + \diag(\theta))z - b + \nu},
\]
where $\rho > 0$ is a penalty parameter we set. Minimizing over each of $z$ and $\theta$ (with the constraint $0 \le \theta \le \tm$) yields the following update rules
\begin{align*}
z^{(k+1)} &= (W^2 + \rho A(\theta^{(k)})^TA(\theta^{(k)}))^{-1}(W^2\hat z + \rho A(\theta^{(k)})^T(b - \nu^{(k)})) \\
\theta^{(k+1)}_i &= S((b_i - a_i^Tz^{(k+1)} - \nu_i^{(k)})/z^{(k+1)}_i, \tm_i), ~~ i=1, \dots, n\\
\nu^{(k+1)} &= \nu^{(k)} + A(\theta^{(k+1)})z^{(k+1)} - b,
\end{align*}
where we have defined $A(\theta) = A + \diag(\theta)$ and $S(x, u) = \min\{\max\{x, 0\}, u\}$ is the clamp function with upper limit, $u$, and we arbitrarily define $S(x/0, u) = 0$ for any $x$, though any value in $[0, u]$ would similarly suffice.

It can be shown that, if a feasible field exists for some $\theta$, then, as $k \to \infty$, the iterates converge to a locally-optimal design $\theta$ and feasible field $z$, for an appropriately large choice of penalty parameter $\rho$~\cite{gao:2018}. In practice, we find that ADMM is fairly robust and converges for a large range of values of $\rho$, though some choices appear to increase convergence speed.

\subsection{Formulations of physical problems}
Here, we describe ways of mapping the photonic inverse design problem into extensions of problem~(\ref{eq:invdes}).

\paragraph{Maxwell's equations in three dimensions.}
Ampere's law and Faraday's law in Maxwell's equations, for a specific frequency $\omega$, can be written as
\begin{align}
\nabla \times H &= - \ii\omega \eps E + J \label{eq:ampere}\\
\nabla \times E &= \ii\omega \mu H\label{eq:faraday},
\end{align}
over some compact region of space $\Omega \subset \reals^3$, with appropriate boundary conditions for $H$ and $E$. Here, $E, H, J: \Omega \to \complex^3$ are the electric field, magnetic field, and the current density, respectively, $\eps, \mu: \Omega \to \reals_+$ are the permittivity and permeability of the space (which we can often control by an appropriate choice of material), respectively. The bold $\ii$---to avoid confusion with the index $i$---is the imaginary unit with $\ii^2 = -1$. We will also assume that we can choose any permittivity and permeability that satisfy $\eps^\mathrm{min}(x) \le \eps(x) \le \eps^\mathrm{max}(x)$ and $\mu^\mathrm{min}(x) \le \mu(x) \le \mu^\mathrm{max}(x)$ at each point of the region $x \in \Omega$.

\subsubsection{Constant permeability}\label{sec:constant-perm}
In many physical design problems, $\mu$ is also a constant that is independent of our material choice (\eg, in the case where we are choosing between silicon or air, under small magnetic field) and constant through space (\ie, $\mu(x) = \mu_c$ for $x \in \Omega$). Assuming this is true, we can write
\[
\nabla \times \nabla \times E = \ii \omega \mu_c \nabla\times H = \omega^2 \mu\eps E + \ii\omega \mu_c J,
\]
by taking the curl of~(\ref{eq:faraday}) and plugging in~(\ref{eq:ampere}). Rearranging gives,
\begin{equation}\label{eq:maxwell}
-\nabla \times \nabla \times E + \omega^2\mu_c \eps E = -\ii\omega\mu_c J.
\end{equation}
All we require is a discretization of $E$, $\eps$, $J$, and the linear operator $-(\nabla \times \nabla \times \cdot)$. There are several standard ways of doing this (\eg, the Yee lattice, see~\cite[\S4.6.4]{chew:1995}), though any method which discretizes the linear operator in the space will suffice. Let $z \in \complex^{3n}$ be the optimization variable corresponding to the discretized field with $z^i \in \complex^n$ being the field along each of the three axes, $i=1, 2, 3$. Then, we can rewrite and discretize~(\ref{eq:maxwell}) as
\[
\big(\underbrace{-\nabla \times \nabla \times}_{A} + \omega^2\mu_c \eps\big) \underbrace{E}_{z} = \underbrace{-\ii\omega\mu_c J}_{b}.
\]
Here, each of $A \in \complex^{3n\times 3n}$ and $b \in \complex^{3n}$ are the corresponding discretizations of the variables they are below.

The next question is: how can we deal with the scalar permittivity term? One simple way is to allow $\theta \in \complex^{3n}$---which roughly corresponds to the discretized version of $\omega^2 \mu_c \eps$---to have a component along each axis, which we will call $\theta^i$ for $i=1, 2, 3$, and to then constrain all axes to be equal---\ie, $\theta^1 = \theta^2 = \theta^3$. Using this idea, we can then write $\diag(\theta) z$, as a discretization of $\omega^2\mu_c \eps E$. Note that, without the equality constraint, $\theta$ would be allowed to vary arbitrarily along each axis.

Finally, we set $\tm$ to be the largest possible value of $\omega^2 \mu_c \eps$ at each point in the discretization (with a similar case for $\theta^\mathrm{min}$), which lets us write the final program as a special case of~(\ref{eq:masks}),
\[
\begin{array}{ll}
\mbox{minimize}   & \frac12 \normsq{W(z - \hat z)} \\
\mbox{subject to} & (A + \diag(\theta))z=b \\
& \theta^1 = \theta^2 = \theta^3,\\
&  \theta^\mathrm{min} \le \theta^i \le \tm, ~~ i=1, 2, 3.
\end{array}
\]

\subsubsection{Arbitrary permeability} In the case where we are also allowed to vary the permeability throughout the space, we can discretize the equations in a similar way. The resulting system will have roughly double the size, but is still---usually, depending on the choice of discretization---relatively sparse.

First, we can write equations~(\ref{eq:ampere}) and~(\ref{eq:faraday}) in the suggestive form
\begin{equation}\label{eq:complete-maxwell}
\begin{bmatrix}
\nabla \times & 0\\
0 & \nabla\times
\end{bmatrix}\begin{bmatrix}
E\\
H
\end{bmatrix} + \ii\omega\begin{bmatrix}
  \eps I & 0 \\
  0 & -\mu I
\end{bmatrix}\begin{bmatrix}
E\\
H
\end{bmatrix} = \begin{bmatrix}
  J\\
  0
\end{bmatrix},
\end{equation}
where $I \in \reals^{3\times 3}$ is the identity matrix. From here, we can perform a similar trick as in~\S\ref{sec:constant-perm}, by rewriting and discretizing~(\ref{eq:complete-maxwell}) in the following way:
\[
\Bigg(-\ii\omega^{-1}\underbrace{\begin{bmatrix}
\nabla \times & 0\\
0 & \nabla\times
\end{bmatrix}}_{A} + \begin{bmatrix}
  \eps I & 0 \\
  0 & -\mu I
\end{bmatrix}\Bigg)\underbrace{\begin{bmatrix}
E\\
H
\end{bmatrix}}_{z} = \underbrace{-\ii \omega^{-1}\begin{bmatrix}
  J\\
  0
\end{bmatrix}}_{b},
\]
where $A \in \complex^{6n \times 6n}$, $z \in \complex^{6n}$, and $b \in \complex^{6n}$ are the discretized versions of the expressions above each. We will write $z_E^i$ for $i=1, 2, 3$ to be the $i$th component of the discretization of the $E$-field, with a similar definition for $z_H^i$.

As before, let $\theta_E^i \in \reals^{n}$ and $\theta_H^i\in\reals^{n}$ be the discretization of the permittivity and permeability, respectively, along each axis $i=1, 2, 3$, with $\theta$ being the concatenation of each component and field over all points in the discretization. To ensure that the permittivity and permeability all remain scalar quantities, we simply constrain each entry of $\theta$  to be equal along all axes at each discretization point, which yields a problem which is a special case of~(\ref{eq:masks}):
\[
\begin{array}{ll}
\mbox{minimize}   & \frac12 \normsq{W(z - \hat z)} \\
\mbox{subject to} & (A + \diag(\theta))z=b \\
& \theta^1_E = \theta^2_E = \theta^3_E,\\
& \theta^1_H = \theta^2_H = \theta^3_H,\\
&  \theta^\mathrm{min}_E \le \theta^i_E \le \tm_E, ~~ i=1, 2, 3,\\
&  \theta^\mathrm{min}_H \le \theta^i_H \le \tm_H, ~~ i=1, 2, 3,
\end{array}
\]
where $\theta^\mathrm{min}_E$ is defined to be the minimum value of $\eps$ at each discretization point with a similar definition for $\theta^\mathrm{max}_E$, $\theta^\mathrm{max}_H$, $\theta^\mathrm{min}_H$.

\end{document}

%% file: defs.tex
\newcommand{\ones}{\mathbf 1}
\newcommand{\reals}{{\mbox{\bf R}}}

\newcommand{\diag}{\mathop{\bf diag}}

\newcommand{\argmin}{\mathop{\rm argmin}}
\newcommand{\argmax}{\mathop{\rm argmax}}

\newcommand{\eg}{{\it e.g.}}
\newcommand{\ie}{{\it i.e.}}

\newcommand{\BEAS}{\begin{eqnarray*}}
\newcommand{\EEAS}{\end{eqnarray*}}
\newcommand{\BEA}{\begin{eqnarray}}
\newcommand{\EEA}{\end{eqnarray}}
\newcommand{\BEQ}{\begin{equation}}
\newcommand{\EEQ}{\end{equation}}
\newcommand{\BIT}{\begin{itemize}}
\newcommand{\EIT}{\end{itemize}}

%% file: comp_imposs_res.bbl
\newcommand{\etalchar}[1]{$^{#1}$}
\begin{thebibliography}{LKBMY13}

\bibitem[AVDB18]{cvxpy:2018}
Akshay Agrawal, Robin Verschueren, Steven Diamond, and Stephen Boyd.
\newblock A rewriting system for convex optimization problems.
\newblock {\em Journal of Control and Decision}, 5(1):42--60, 2018.

\bibitem[BEKS17]{bezanson:2017}
Jeff Bezanson, Alan Edelman, Stefan Karpinski, and Viral~B. Shah.
\newblock Julia: A fresh approach to numerical computing.
\newblock {\em SIAM review}, 59(1):65--98, 2017.

\bibitem[BPC{\etalchar{+}}11a]{boyd:2011}
Stephen Boyd, Neal Parikh, Eric Chu, Borja Peleato, Jonathan Eckstein, et~al.
\newblock Distributed optimization and statistical learning via the alternating
  direction method of multipliers.
\newblock {\em Foundations and Trends in Machine learning}, 3(1):1--122, 2011.

\bibitem[BPC{\etalchar{+}}11b]{admm:2011}
Stephen Boyd, Neal Parikh, Eric Chu, Borja Peleato, Jonathan Eckstein, et~al.
\newblock Distributed optimization and statistical learning via the alternating
  direction method of multipliers.
\newblock {\em Foundations and Trends{\textregistered} in Machine learning},
  3(1):1--122, 2011.

\bibitem[BV04]{cvxbook}
Stephen Boyd and Lieven Vandenberghe.
\newblock {\em Convex optimization}.
\newblock Cambridge university press, 2004.

\bibitem[BW13]{born:2013}
Max Born and Emil Wolf.
\newblock {\em Principles of optics: electromagnetic theory of propagation,
  interference and diffraction of light}.
\newblock Elsevier, 2013.

\bibitem[Che95]{chew:1995}
Weng~C. Chew.
\newblock {\em Waves and fields in inhomogeneous media}.
\newblock IEEE press, 1995.

\bibitem[DCB13]{ecos:2013}
Alexander Domahidi, Eric Chu, and Stephen Boyd.
\newblock {ECOS}: An {SOCP} solver for embedded systems.
\newblock In {\em Control Conference (ECC), 2013 European}, pages 3071--3076.
  IEEE, 2013.

\bibitem[DHL17]{dunning:2017}
Iain Dunning, Joey Huchette, and Miles Lubin.
\newblock Ju{MP}: A modeling language for mathematical optimization.
\newblock {\em SIAM Review}, 59(2):295--320, 2017.

\bibitem[Fer36]{fermi:1936}
Enrico Fermi.
\newblock {\em Thermodynamics}.
\newblock Snowball Publishing, 1936.

\bibitem[GGC18]{gao:2018}
Wenbo Gao, Donald Goldfarb, and Frank~E. Curtis.
\newblock {ADMM} for multiaffine constrained optimization.
\newblock {\em arXiv preprint arXiv:1802.09592}, 2018.

\bibitem[GO18]{gurobi:2018}
LLC Gurobi~Optimization.
\newblock Gurobi optimizer reference manual, 2018.

\bibitem[HG12]{haftka:2012}
Raphael~T. Haftka and Zafer G{\"u}rdal.
\newblock {\em Elements of structural optimization}, volume~11.
\newblock Springer Science \& Business Media, 2012.

\bibitem[LKBMY13]{lalau:2013}
Christopher~M. Lalau-Keraly, Samarth Bhargava, Owen~D. Miller, and Eli
  Yablonovitch.
\newblock Adjoint shape optimization applied to electromagnetic design.
\newblock {\em Optics express}, 21(18):21693--21701, 2013.

\bibitem[LV10]{lu:2010}
Jesse Lu and Jelena Vu{\v{c}}kovi{\'c}.
\newblock Inverse design of nanophotonic structures using complementary convex
  optimization.
\newblock {\em Optics express}, 18(4):3793--3804, 2010.

\bibitem[LV13]{lu:2013}
Jesse Lu and Jelena Vu{\v{c}}kovi{\'c}.
\newblock Nanophotonic computational design.
\newblock {\em Optics express}, 21(11):13351--13367, 2013.

\bibitem[LVBL98]{lobo:1998}
Miguel~S. Lobo, Lieven Vandenberghe, Stephen Boyd, and Herv{\'e} Lebret.
\newblock Applications of second-order cone programming.
\newblock {\em Linear algebra and its applications}, 284(1-3):193--228, 1998.

\bibitem[MLP{\etalchar{+}}18]{molesky:2018}
Sean Molesky, Zin Lin, Alexander~Y. Piggott, Weiliang Jin, Jelena Vu{\v
  c}kovi{\'c}, and Alejandro~W. Rodriguez.
\newblock Inverse design in nanophotonics.
\newblock {\em Nature Photonics}, 12(11):659, 2018.

\bibitem[MPR{\etalchar{+}}16]{miller:2016}
Owen~D. Miller, Athanasios~G. Polimeridis, M.T.~Homer Reid, Chia~Wei Hsu,
  Brendan~G. DeLacy, John~D. Joannopoulos, Marin Solja{\v{c}}i{\'c}, and
  Steven~G. Johnson.
\newblock Fundamental limits to optical response in absorptive systems.
\newblock {\em Optics express}, 24(4):3329--3364, 2016.

\bibitem[NUS{\etalchar{+}}10]{noreland:2010}
Daniel Noreland, Rajitha Udawalpola, Pablo Seoane, Eddie Wadbro, and Martin
  Berggren.
\newblock An efficient loudspeaker horn designed by numerical optimization: an
  experimental study.
\newblock {\em Report UMINF}, 10, 2010.

\bibitem[OCPB16a]{scs_paper:2016}
Brendan O’Donoghue, Eric Chu, Neal Parikh, and Stephen Boyd.
\newblock Conic optimization via operator splitting and homogeneous self-dual
  embedding.
\newblock {\em Journal of Optimization Theory and Applications},
  169(3):1042--1068, 2016.

\bibitem[OCPB16b]{scs:2016}
Brendan O’Donoghue, Eric Chu, Neal Parikh, and Stephen Boyd.
\newblock {SCS}: Splitting conic solver, version 1.2. 6, 2016.

\bibitem[Pur77]{purcell:1977}
Edward~M. Purcell.
\newblock Life at low {R}eynolds number.
\newblock {\em American journal of physics}, 45(1):3--11, 1977.

\bibitem[SFJM18]{shim:2018}
Hyungki Shim, Lingling Fan, Steven~G. Johnson, and Owen~D. Miller.
\newblock Fundamental limits to near-field optical response, over any
  bandwidth.
\newblock {\em arXiv preprint arXiv:1805.02140}, 2018.

\end{thebibliography}
